\documentclass[twocolumn ,showpacs,preprintnumbers,amsmath,amssymb]{revtex4}

\usepackage{graphicx}
\usepackage{dcolumn}
\usepackage{bm}

\preprint{VERSION-10}

\begin{document}

\title{Complex magnetic phase diagram of $\rm CePd_{2-x}Ni_xAl_3$
around its critical concentration}

\author{J. G. Sereni}
    \email{jsereni@cab.cnea.gov.ar}
   \homepage{http://cabbat1.cnea.gov.ar/~jsereni}
\author{P. Pedrazzini}
\affiliation{Lab.\ Bajas Temperaturas, Centro At\'omico Bariloche
(CNEA) 8400 S.C. de Bariloche, Argentina}
\author{E. Bauer}
\author{A. Galatanu}
\affiliation{Institut f\"ur Festk\"orperphysik, Technische
Universit\"at Wien, A-1040 Wien, Austria}
\author{Y. Aoki}
\author{H. Sato}
\affiliation{Department of Physics, Tokyo Metropolitan University,
Minami-Ohsawa 1-1, Hachioji-Shi, Tokyo 192-0397, Japan}

\date{\today}

\begin{abstract}

Thermal, magnetic and transport measurements are reported on
$\rm CePd_{2-x}Ni_xAl_3$ in the $0\leq x \leq 1$ range,
including the effect of pressure ($p$) and magnetic
field on some selected samples. The low temperature results indicate
that long range antiferromagnetic order is stable up to $x=0.2$,
while between 0.25 and 0.5 magnetic disorder gives rise to
non-Fermi-liquid (NFL) behavior. In this region, the low
temperature specific heat can be described by the sum of two
components, the major one showing a $C_p/T=\gamma_0 -\gamma_1 \sqrt{T}$
dependence, while the minor one includes a decreasing number of degrees
of freedom related to a residue of short range order. The latter
extrapolates to zero between $0.45<x_{cr}<0.5$. Electrical
resistivity ($\rho$) studies performed under pressure for $x=0.5$
allow to investigate the evolution of the NFL state right
above the critical point, where the exponent of  $\rho \propto T^n$
is found to increase from $n=1$ (for $p=0$) up to $n=2$ (for
$p=12$~kbar). The latter is observed for $x=1.0$ already at normal
pressure, indicating the onset of the Fermi Liquid behavior. Doping
and pressure effects are compared by fitting high temperature
resistivity data employing a unique function which allows to
describe the evolution of the characteristic energy of this series
through a large range of concentration and pressure.
\newline

\end{abstract}

\maketitle

\section{Introduction}

Numerous magnetic phase diagrams of Ce-lattice systems have been investigated
in recent years motivated by novel physical properties observed around
their magnetic instability region. In these systems the variation of the
ordering temperature ($T_{\rm N}$) is usually driven by controlling parameters
like pressure ($p$) or Ce-ligand substitution ($x$). At low temperatures, the
divergences observed in their thermodynamical parameters correspond to a
non-Fermi-liquid (NFL) behavior \cite{1HVL,2Stew}, indicating the
vicinity to a quantum critical point (QCP) at $T=0$ \cite{3StaBarb}.
This is a consequence of substantial modifications in the nature
of the fluctuations related to the magnetic phase transition. The possibility
of experimental access to this new regime, where quantum fluctuations
compete in energy with the thermal ones, has concentrated a significant effort
on the study of magnetic phases with such characteristics.

Because the QCP occurs at zero temperature, it is frequently found that some
magnetic phase diagrams include na\"ive extrapolations for
$T_{\rm N}(x,p)\to 0$. A
strict analysis of the experimental results indicates that only in a few
Ce-lattice systems $T_{\rm N}(x,p)$ can be unambiguously traced down to very
low
temperature, i.e. at least down to $T_{\rm N}(x,p)/T_{\rm N}(0,0)<0.1$ like in
$\rm CeCu_{6-x}Au_x$ \cite{1HVL}
or $\rm CeIn_{3-x}Sn_x$ \cite{44Pedra}. Furthermore,
an ample comparison of the magnetic phase diagrams performed on a large number
of Ce-systems have shown that a clear distinction can be made among different
types of phase boundaries. These different behaviors can be sorted in three
groups distinguished by the following characteristics \cite{5JPSJ}: i) the
already quoted systems with $T_{\rm N}(x,p)\to 0$; ii) those whose phase
boundaries
vanish at finite temperature ($T_{\rm N}(x,p) / T_{\rm N}(0,0)\geq 0.4$) and
iii) the
systems where $T_{\rm N}$ is weakly doping or pressure dependent before
vanishing.

In the scope of Doniach´s model \cite{6Doni}, the main properties of the
magnetic
phase diagram are accounted for by two related parameters: $T_{\rm N}$ and
$T_K$,
which depend on the same coupling parameter $J_{ex}$ between local moments
and conduction electron spins.
Nevertheless, different doping and pressure dependencies were
experimentally observed for $T_{\rm N}$ and T$_K$ in each of the
groups mentioned \cite{7Ser01}.
For example, in the systems included in case ii) $T_{\rm N}(x,p)$ decreases
while the temperature scale characterizing the Kondo
lattice $T_0$ ($\propto T_K$) practically does not change.
On the contrary, those included in case
iii) show a weak doping or pressure dependence of $T_{\rm N}$, while
$T_{\rm 0}$ increases monotonously.
It is therefore of great interest to perform detailed experimental
investigations on such Ce-systems in order to unambiguously determine
the $T_{\rm 0}(x,p)$ behavior throughout the transition between their magnetic
and non-magnetic phases.

From previous studies on its physical properties,
$\rm CePd_{2-x}Ni_xAl_3$ \cite{8Kita,14Tang}
can be considered as an exemplary candidate
for this investigation. By doping Pd sites with smaller Ni atoms
it is possible to induce a continuous change from antiferromagnetic
$\rm CePd_2Al_3$ (T$_{\rm N}=2.8$~K \cite{8Kita}) to non-magnetic
$\rm CeNi_2Al_3$
\cite{9Fuji}.
In this system $T_{\rm N}(x,p)$ and the related
specific heat jump ($\Delta C_{\rm m}$) are weakly affected by Ni doping
up to $x=0.2$ \cite{10Galanta,11Bauer,12Isika} and pressure \cite{13Eichler}.
Above that concentration, $\Delta C_{\rm m}$ transforms
into a broad anomaly \cite{12Isika}. An advantage of this system
is that doping effects can be compared with pressure studies performed on
stoichiometric $\rm CePd_2Al_3$ \cite{13Hauser,14Tang},
where $T_{\rm N}$ vanishes at $\approx 12$~kbar.
Keeping in mind these features, we have performed
a more detailed and comprehensive investigation on
thermal, magnetic and transport
properties, including for such a purpose new samples with its Ni content
tuned to the critical concentration.
This investigation also includes
electrical resistivity measurements under pressure
and magnetic field on some selected concentrations.
This procedure allows a quantitative comparison between alloying,
pressure and field effects.
All together, this large amount of information permits to access to a
unified description of this system which presents a rich phase diagram
around its QCP and to compute the evolution of the characteristic energy
scale of this system over more than one decade.

\section{Experimental and results}

Samples with Ni concentrations ranging between 0.03$<x<$1 were prepared by
melting appropriate amounts of elements using a high frequency melting
procedure and
subsequent heat treated at T=900$^o$ C during two weeks \cite{10Galanta}. No
foreign phases were detected from X-ray diffraction patterns obtained
applying Co$K_{\alpha}$ radiation. Specific heat measurements on samples of
about 1gr. were performed in a semi-adiabatic calorimeter at temperatures
ranging from 0.2K up to 30K, using a heat pulse technique. A standard SQUID
magnetometer served for the determination of the magnetization from 2K up to
room temperature in a 1Tesla magnetic field. The ac-susceptibility of one of
the
samples was measured using the mutual inductance technique with a lock-in
amplifier as detector working at 12.8 kHz with an excitation amplitude of
$\approx 10 \mu T$. The electrical resistivity and magnetoresistivity were
measured using a four probe dc-method in the temperature range from 0.5K
up to room temperature and fields up to 12Tesla. A piston-cylinder
pressure cell with a paraffin
mixture as pressure transmitter served to generate hydrostatic
pressure up to about 12kbar. The absolute value of the pressure was determined
from the superconducting transition temperature of lead.

The evolution of the structural parameters with increasing Ni concentration is
shown in
Fig.1. As expected from the relative atomic volumes of the pure elements
(Ni and Pd), the molar volume (V$_m$) of CePd$_{2-x}$Ni$_x$Al$_3$ decreases
with
increasing Ni content at a rate:
$\Delta V/V_m*1/\Delta x=dlnV/dx=0.033(2)/$Ni at. (see Fig.1a).
Also the ``c/a`` ratio between the lattice parameters of this hexagonal system
decreases by about 1.3\% per Ni atom, producing a slight modification in the
relative positions of the Ce-neighboring atoms. For a more detailed analysis,
we have
evaluated the corresponding
interatomic distances as: d$_{Ce-TM}=a \sqrt{3}$ and
d$_{Ce-Al}=1/2\sqrt{a^2+c^2}$, where TM indicates Pd or Ni.
Those distances are compared in Fig.1b with
the respective atomic radius ($r_{Ce^{3+}}=1.86\AA$, $r_{Pd}=1.37\AA$,
$r_{Ni}=1.25\AA$ and $r_{Al}=1.43\AA$ \cite{15radius}) computing
an effective interatomic spacing as:
$\Delta_{Ce-Z}=d_{Ce-Z}-(r_{Ce^{3+}}+r_{Z})$ (with $Z$=TM or Al). In a simple
rigid-spheres picture, $\Delta_{Ce-Z}<0$ indicates
a reduction of the Cerium Wigner-Seitz cell with respect to that of pure
metal.
We notice that this reduction occurs for $\Delta_{Ce-TM}$ but not for
$\Delta_{Ce-Al}$.
Nevertheless, despite the reduction of $V_m(x)$,
the positive slope of $\Delta_{Ce-TM}(x)$ indicates that the Ce-$TM$ overlap
becomes weaker with Ni doping.
On the contrary, $\Delta_{Ce-Al}$ decreases with doping,
although no direct atomic overlap between
Ce and Al is expected at any concentration.

\begin{figure}
\begin{center}
\includegraphics[angle=0,width=.40\textwidth] {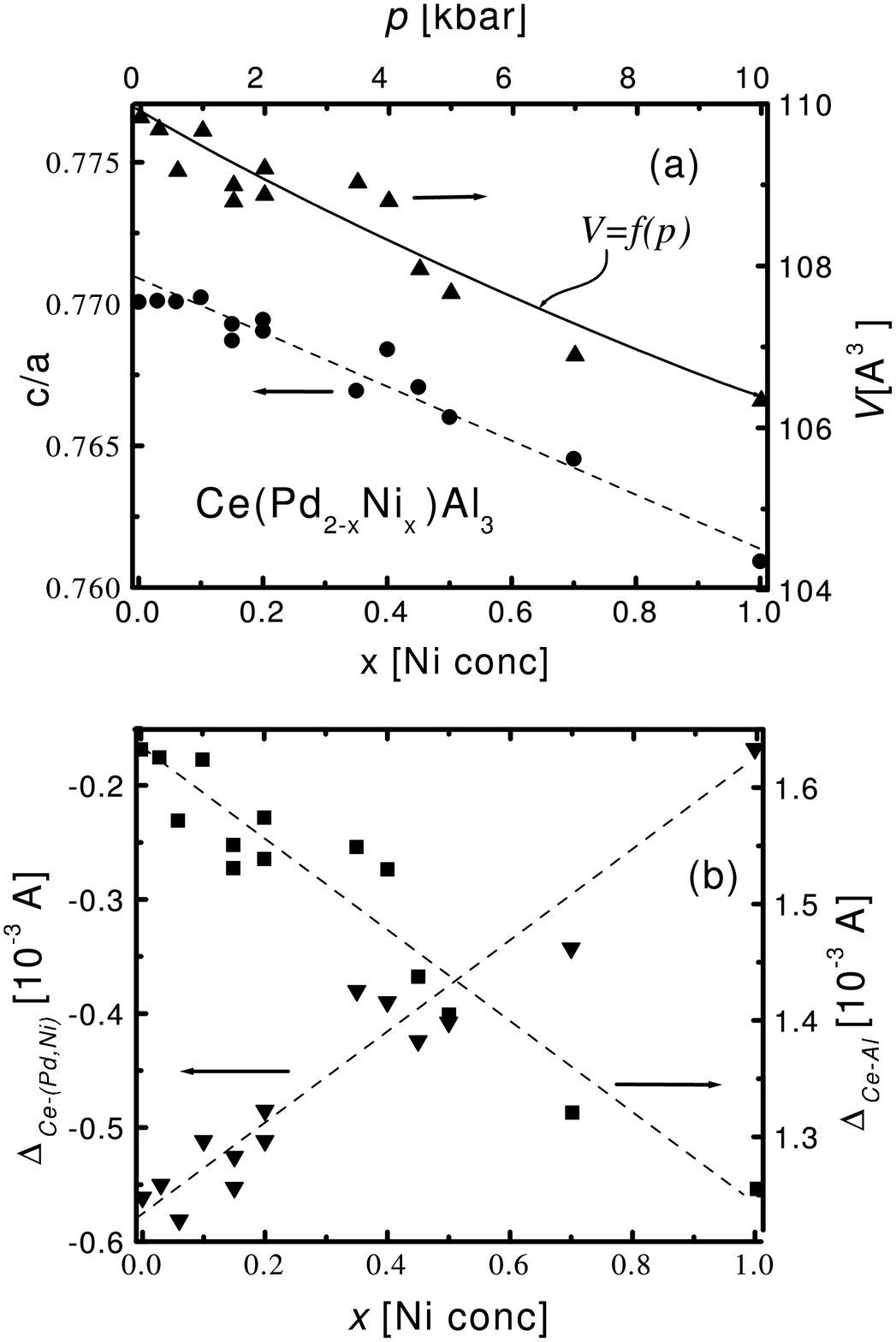}
\end{center}
\caption{(a)
Lattice parameters ratio $c/a$ (left side) and unit cell volume $V$
(right side) as a function of Ni concentration. The solid line indicates
the volume (right side) dependence on pressure (top axis) of
CePd$_2$Al$_3$, after \cite{14Tang}.
(b) Comparison of the effective interatomic spacings respect to
those of pure elements (see the text).}
\label{F1latpar}
\end{figure}

The magnetic component of the specific heat ($C_{\rm m}$) is obtained by
subtracting the phonon contribution ($C_{\rm {ph}}$) from the measured values:
$C_{\rm m} = C _{\rm P} - C_{\rm {ph}}$, where $C_{\rm {ph}}$ was taken from
the reference compound LaPd$_2$Al$_3$.
Data for the samples with $x \leq 0.20$ are displayed in Fig.2(a) using a
 a $C_{\rm m}/T$ vs $T^2$ representation.
The observed $C_{\rm m}/T$=$\gamma$+B$T^2$ dependence for $T<T_{\rm N}$
corresponds to
long range antiferromagnetic (AF) order, where B $\propto J_{ex}^{-3}$ \cite{
16Gopal}.
Because this temperature dependence is
found to be almost independent of concentration (B=0.11$\pm$0.01 J/molK$^4$),
one concludes that $J_{ex}$ has no significant variation up to $x$=0.20.
This interpretation is consistent with insignificant change in $T_K^{GS}$
as will be shown in Fig.8.
Above that concentration a clear change of $C_{\rm m}(T)$ is observed at
$T\leq T_{\rm N}$ (see Fig.2b).
The $C_{\rm m}/T$ peak at $T_{\rm N}$ transforms into a broad anomaly,
whose maxima show a minimum value at $x$=0.3. Additionally, the
temperatures of the maxima extrapolate to zero at $x_{cr}\approx$ 0.45.
At higher Ni concentration no magnetic transition is detected down to 0.4K.

\begin{figure}
\begin{center}
\includegraphics[angle=0,width=.47\textwidth] {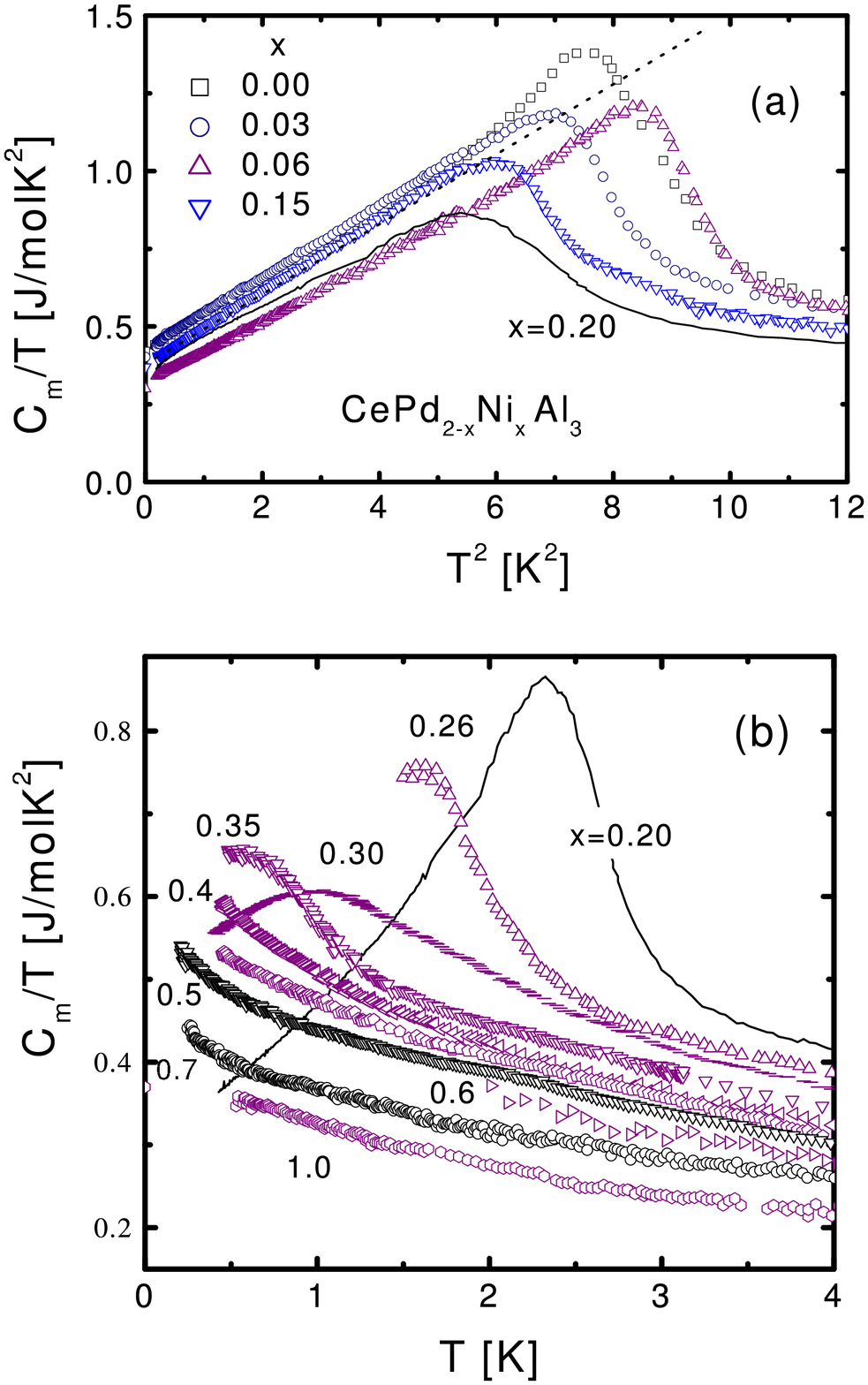}
\end{center}
\caption{Magnetic contribution to the specific heat $C_{\rm m}$ divided by T
as:
(a) $C_{\rm m}/T$ vs. $T^2$ representation for the samples with $x\leq 0.20$.
The dotted line represents a $\gamma$+B$T^2$ function (see the text).
(b) linear temperature dependence for the $x\geq 0.2$ samples to show the
$C_{\rm m}/T$ vs. $T$ evolution around the critical concentration. Results for
$x$=
0.26 and 0.6 are taken from Ref. \cite{12Isika}.}
\label{F2TTySG}
\end{figure}

The temperature dependent magnetic susceptibility, measured
for samples with $x\leq$0.45 is shown in Fig.3 in a
semi-logarithmic representation. The results are compared with those
reported in the literature for $x$=0.26 \cite{12Isika}, $x$=1
\cite{10Galanta} and $x$=2 \cite{9Fuji}. While at high temperature
nearly the full magnetic moment ($\mu_{eff}=2.54\mu_B$) of the Ce ion
is observed, a deviation from the Curie-Weiss law at low temperature
indicates a weakening of the effective moment owing the reduction of
the thermal population of excited CF levels. A well defined maximum at
$T_{\rm N}$ is evident for samples exhibiting long
range magnetic order (LRMO). On the contrary, for samples with $x > 0.2$ there
is a
tendency to saturation at low temperature with a magnetization
(evaluated at T=2K) dropping from 0.05 to 0.028 emu/mol between
$x$=0.15 and 0.45. Despite this reduction, an
ac-susceptibility measurement performed on the $x$=0.35
sample down to 0.5K indicates a remainder magnetic anomaly at 0.7K, in
agreement
with $C_{\rm m}/T$ results (see inset of Fig.3).

The absolute value of the Curie-Weiss temperature $\theta_W$,
extrapolated from high temperature (200$<T<$300K),
increases monotonously with Ni concentration
from $\theta_W$(0$<x<$0.15)$\approx-$40K up to $\theta_W$($x$=1)$\approx-$100
K.
This concentration dependent variation of $\theta_W$ results from
two competing contributions, one originated from the antiferromagnetic
molecular field and the other arises from the increase of the Kondo
temperature. While the former is expected to decrease as Ce magnetic moments
weaken, the latter increases with $T_K$. Therefore, the second is
expected to dominate the value of $\theta_W$ once $T_N<<\theta_W$.

\begin{figure}
\begin{center}
\includegraphics[angle=270,width=.47\textwidth] {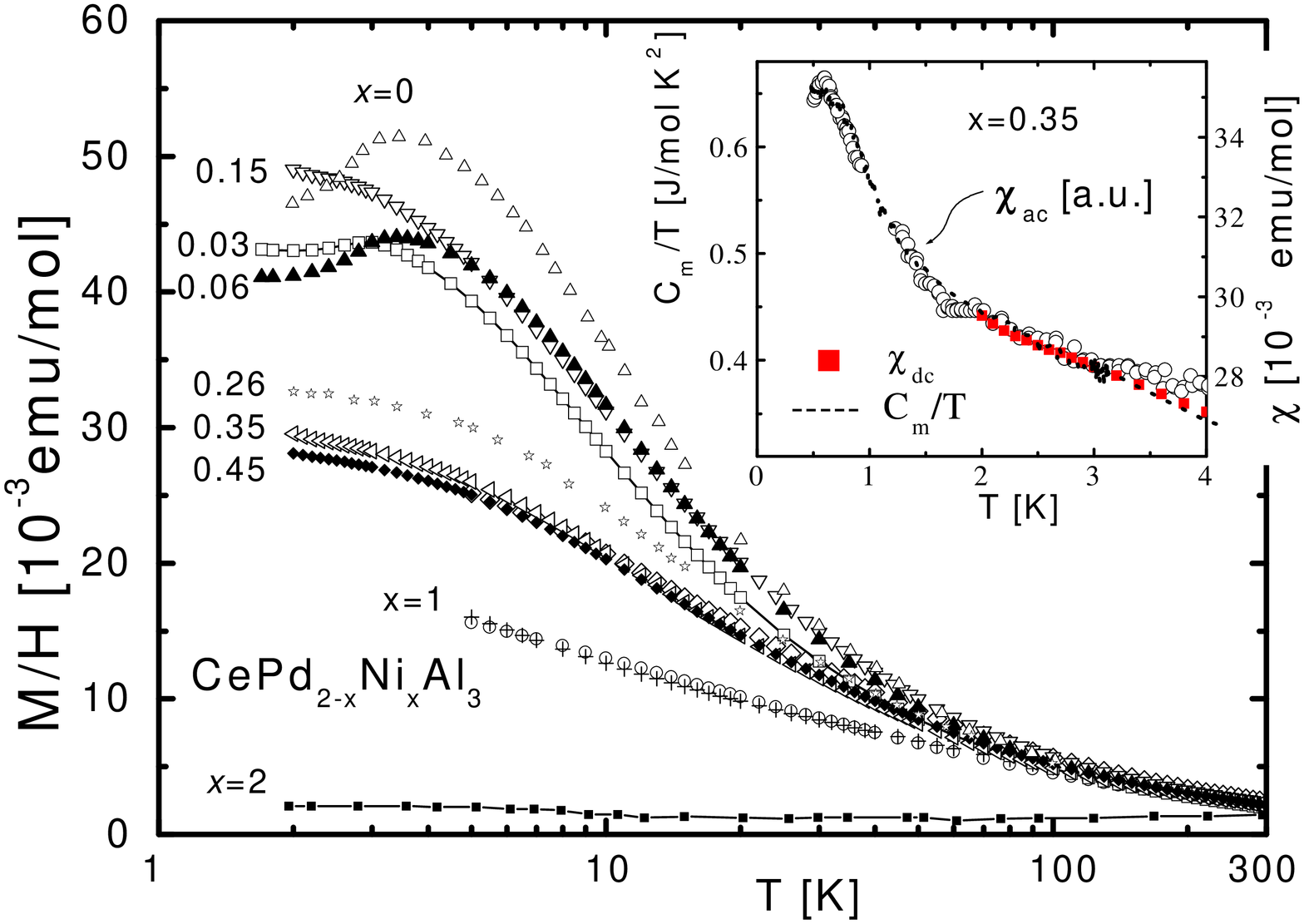}
\end{center}
\caption{Temperature dependent magnetization divided by applied field
(1 Tesla), in a semi-logarithmic representation. Results for $x$=0.26
\cite{12Isika}, 1.0 \cite{10Galanta} and 2.0 \cite{9Fuji} are
included for comparison. Inset: ac-susceptibility ($\bigcirc$)
measurements performed on sample $x$= 0.35 down to 0.5K, compared
with specific heat (dashed) and dc-magnetization ($\Box$) results.}
\label{F3susdcyac}
\end{figure}

The electrical resistivity of CePd$_{2-x}$Ni$_x$Al$_3$ was investigated as
a function of Ni concentration, hydrostatic pressure and magnetic field,
providing valuable information on the evolution of the
ground state, undergoing a substitution driven magnetic to non-magnetic
transformation.
In order to evaluate the role of the Kondo effect in the AF phase,
we have studied the field dependent resistance $R(H,T)$ of the sample
$x=0.15$ up to H=12T.
Experimental values obtained from isotherms within
$0.5\leq T \leq 20K$ are shown in Fig.4a as $\Delta R /R$ vs. $H$.
$\Delta R /R = R(0) - R (H) /R(0)$, where $R(0)$ and $R(H)$ are the
resistances
without and with magnetic field, respectively.
Above $T_{\rm N}$, the variation of $\Delta R /R$ can be scaled by a reduced
magnetic field ($H/H^*$) as predicted for Kondo systems \cite{17Schlot},
which in this case is found to be $H^*=(1+1.5T)k_{\rm B}/\mu_{\rm {eff}}$
(see fig.4b).
As expected, deviations of that scaling are observed for $T<T_{\rm N}$ (solid
curves).
Similar scaling (not shown) is obtained for $x$=0.06.

Around the critical concentration, the NFL behavior is identified
by a power law dependence of the electrical resistivity
$R(T)\propto T^n$. In order to better establish the actual
extension of the NFL region on the non-magnetic side of the
critical region, we have analyzed the $R(T)$ dependence of the
$x$=0.2 and $x$=0.5 samples up to $p$=12kbar \cite{10Galanta}, the
latter displayed in an $R$ vs $T^n$ representation in Fig.4c. The
exponent $n(x,p)$ is extracted as the one that better defines a
straight line at low temperature. Because of the proximity of
these samples to the critical point, there is a significant
variation of the exponent from $n(x$=0.2,$p$=0)=0.9 up to
$n(x$=0.5,$p$=11kbar)=2, expected for a Fermi liquid (FL) system.
Similar procedure was applied to the $x$=1 sample, which behaves
as a FL at normal pressure. Despite the narrow temperature range
from which these exponents are extracted, for our purpose the
relevant information concerns the $n(x,p)$ evolution rather than
its actual absolute value, as shown in Fig.7.

\begin{figure}
\begin{center}
\includegraphics[angle=0,width=.47\textwidth] {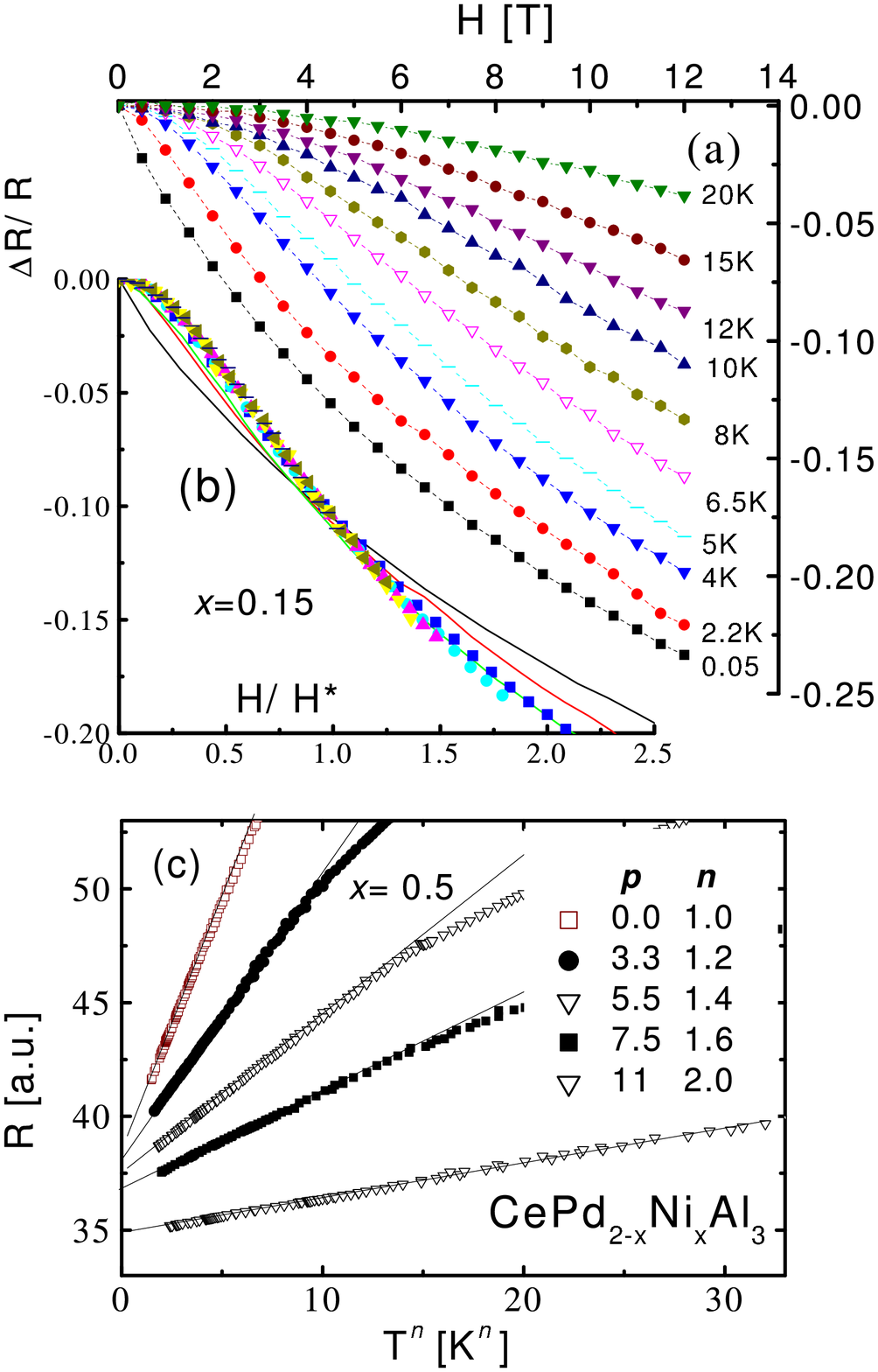}
\end{center}
\caption{(a) Isothermal magnetoresistance $\Delta R/R$ of
$x$=0.15 at different temperatures. (b) Magnetoresistance
as a function of the normalized magnetic field H/H$^*$
(symbols for $T>T_{\rm N}$ and lines for $T\leq T_{\rm N}$).
(c) Low temperature resistance plotted as $R(T)$ vs. $T^n$ dependence for
$x$=0.5 where $n$ is a function of pressure ($p$[kbar]).}
\label{F4magnetor}
\end{figure}

High temperature resistivity measurements were performed on samples ranging
between $x$= 0.1 and 1, under pressures up to 12kbar. In Fig.5 we
compare the $R(T)$ dependence of some selected samples ($x$=0.2; 0.3;
0.5 \cite{10Galanta} and 1.0) with the $\rho(T)$ of the
stoichiometric compound CePd$_2$Al$_3$ \cite{13Hauser} at different
values of applied pressures ($p$= 0; 7; 15; 27; 53 and 63kbar).
For clarity, only those $R(T,p)$ measurements performed on the
stoichiometric compounds which coincide with those obtained on different
Ni alloys (at normal or under pressure), are included in the figure.
They are: for $x$=0.2 at 7kbar and for $x$=0.3 at 15kbar.
In the case of $x$=0.5, the best matching is observed selecting the
measurement
done under 3.3kbar and that of CePd$_2$Al$_3$ under 27kbar. In the
case of $x$=1, the measurements performed under 3.3 and 11kbar are
compared with those at 53 and 63kbar on CePd$_2$Al$_3$, respectively.
In order to scale doping and pressure effects in Fig.5, the
respective $R(T)$ values are adjusted and the high temperature slopes
normalized. The electrical resistivity of the reference compound
LaPd$_2$Al$_3$ is also included to show that at room temperature the
slope of $\rho(T)$ of the Ce samples primarily derives from the phonon
contribution. The temperature of the maximum of $R(T)$
($T^{\rho}_{max}$) is also depicted, showing an increase
with Ni doping from about 28K in CePd$_2$Al$_3$ up to above room temperature
in CePdNiAl$_3$.
Already at $x$=0.5 that maximum becomes so wide that it
mixes with the phonon contribution, impeding any precise
determination.

\begin{figure}
\begin{center}
\includegraphics[angle=0,width=.47\textwidth] {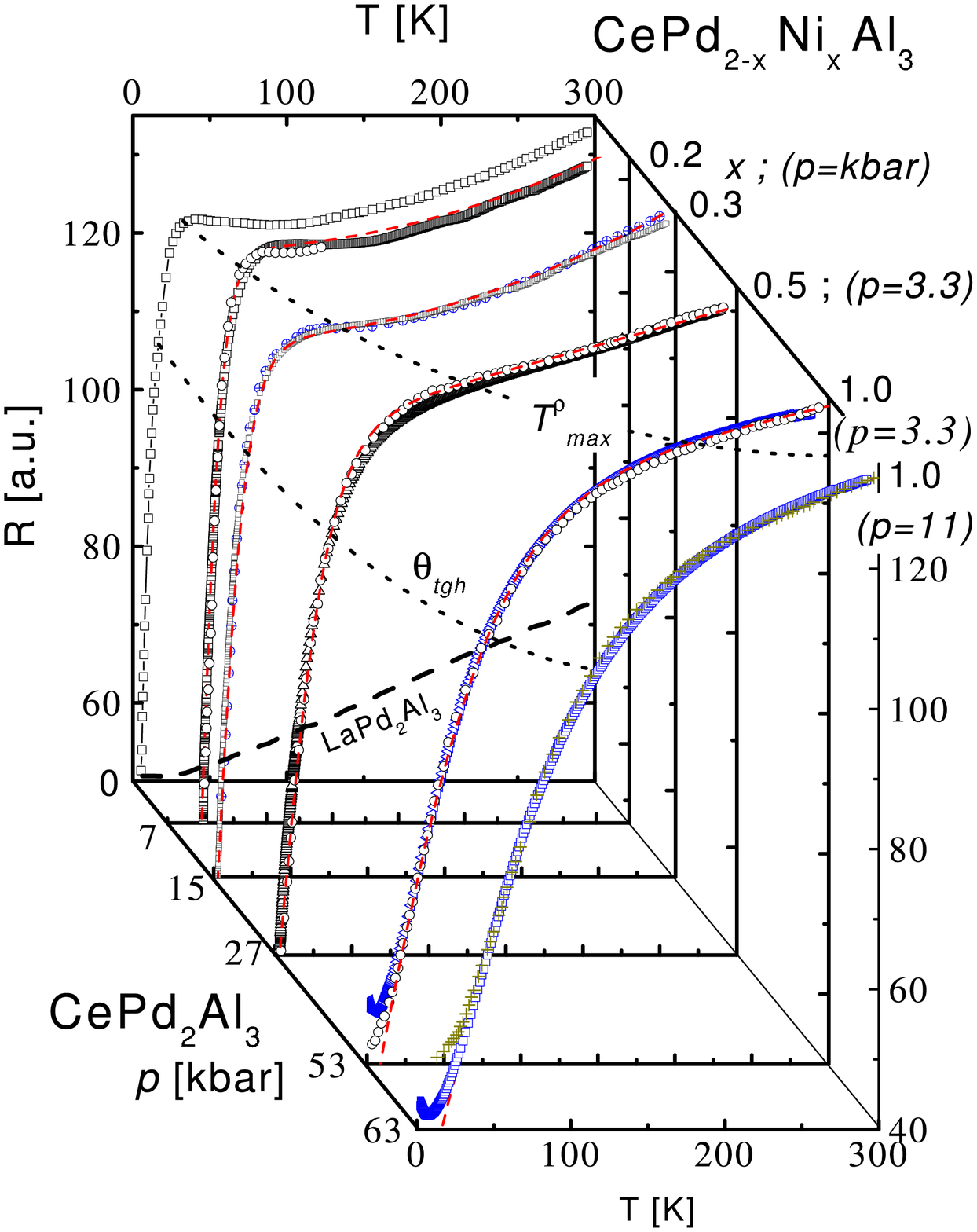}
\end{center}
\caption{Temperature dependent electrical resistance of some selected Ni-doped
samples: $x$=0.2; 0.3; 0.5 (at 3.3kbar) and
1.0 (at 3.3 and 11kbar) (top right axis), in comparison with the resistivity
of stoichiometric CePd$_2$Al$_3$ \cite{13Hauser} at different pressures:
$p$= 0; 7; 15; 27; 53 and 63kbar (bottom left axis).
LaPd$_2$Al$_3$ is included as reference to show the phonon contribution.
$T^{\rho}_{max}$ (dotted curve) indicates the concentration dependent
evolution
of the temperature of the resistivity maximum and $\theta_{\rm {tgh}}$ (dashed
curve)
is a generalized parameter used for comparison (see the text).}
\label{F5RsNivsPrs}
\end{figure}

\section{Discussion}

The specific heat jump at $T_{\rm N}$ allows
to determine the magnetic phase boundary as a
function of concentration up to $x$=0.20. Within that range of concentration,
the well defined jump is followed at lower temperature by
a C$_m(T)\propto T^3$ dependence, which characterizes
a stable AF phase \cite{16Gopal}.
Nevertheless, in this system the LRMO collapses at
$x\geq$ 0.26, where short range (SR) magnetic interactions take over.
As mentioned before, the maximum value of $C_{\rm m}/T$ decreases up to $x$=
0.3 where that tendency reverses. 
Because of the proximity to the critical point, we have
performed a detailed analysis of the $C_{\rm m}(T)/T$ dependencies of the
alloys included within the 0.3$\leq x \leq$ 0.5 range.
As shown in Fig.6a, the measurements are well
described by an expression accounting for two contributions:
$C_{\rm m}(T)/T=C_{\rm {NFL}}/T+C_{\rm {SR}}/T$,
where $C_{\rm {NFL}}/T=\gamma_0 (1-\sqrt{T/D})$
\cite{18Aoki}, and $C_{\rm {SR}}/T$ accounts for the short range interactions.
As a result of this description one finds that there is a majority fraction
of degrees of freedom corresponding to the NFL component (dotted line),
while the entropy related to
$C_{SR}/T$ extrapolates to zero for $x$=0.45, in coincidence with the
evolution of the temperature of the maximum of $C_{\rm m}(T)/T$ (see Fig. 6a).
The ac-susceptibility measurement performed for $x$=0.35 (shown in the
inset of Fig. 3) confirms the anomaly in the specific heat as due to a weak
magnetic transition.

\begin{figure}
\begin{center}
\includegraphics[angle=0,width=.47\textwidth] {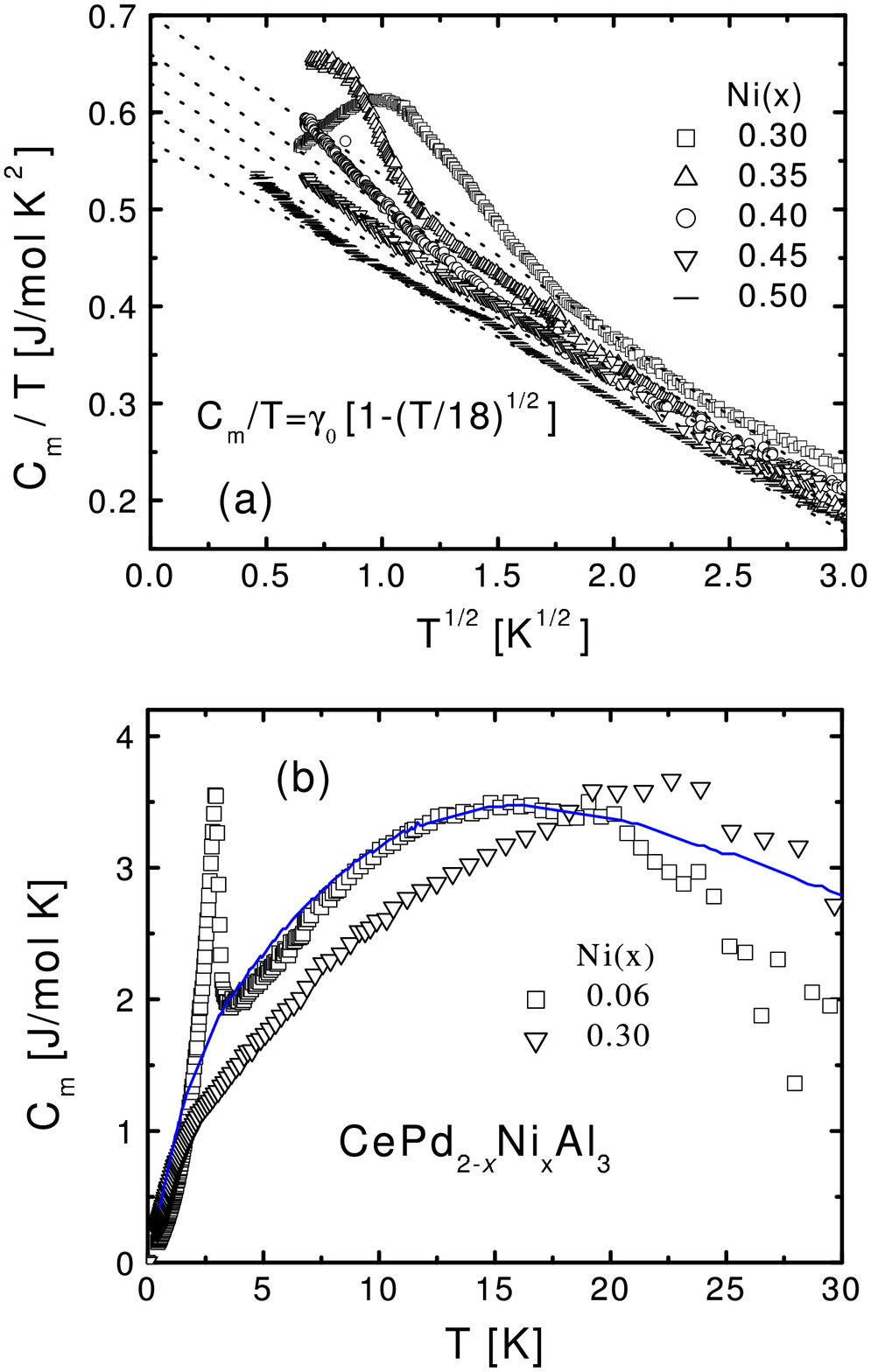}
\end{center}
\caption{(a) Magnetic contribution to specific heat $C_{\rm m}$ plotted as
$C_{\rm m}/T$ vs. $\sqrt(T)$ for the 0.3$\leq x \leq $0.45 samples,
to make evident the $C_{\rm {NFL}}/T=\gamma_0 (x) (1-\sqrt{T/D})$
and $C_{\rm {SR}}/T$ contributions. Dotted lines are extrapolations to T=0
to extract $\gamma_0 (x)$.
(b) Temperature dependent magnetic specific heat up to 30K to show the
contribution of the first excited CF doublet.
Solid line is a fit according to the model of Desgranges \& Rasul
\cite{19DesgrRasul}.}
\label{F6}
\end{figure}

Above $x$=0.45, the changes in the magnetic ground state can be traced
following the evolution of exponent "n" in the $R \propto T^n$ dependence, as
already quoted for the sample $x$=0.5 in Fig. 4c. Its variation between $1\leq
n \leq 2$ (for 0$\leq p \leq$11kbar) is an indication that the NFL behavior
dominates the low temperature physical properties even behind the critical point.
The same procedure
was applied to $x$=1.0, where the exponent $n$=2 marks the entrance to the
Fermi-liquid regime. On the other hand, measurements performed on the sample
$x$=0.2 evidence a tuning towards the critical point at p=8kbar.

We have collected these results in a low temperature
phase diagram shown in Fig.7, combining Ni concentration with pressure
dependencies of $R(T)$ measured on samples with $x$=0.2, 0.5 \cite{10Galanta}
and 1.0.
This analysis shows the complex behavior of this system around its critical
point. Between the LRMO ($x\leq 0.2$) and the FL ($x\geq 1.0$) phases there is
a region (0.2 $\leq x \leq$ 0.5) where the growing NFL component coexists with
an
exhausting fraction of SR interactions. The QCP is then found around
0.45 $\leq x \leq$ 0.5 by both $C_{\rm m}/T=\gamma_0 (1-\sqrt{T/D})$
and $R \propto T^n$. Noteworthy is the fact that approaching the QCP
(i.e. between $x$=0.3 and 0.5) the factor D is practically constant (D
$\approx$18K)
while $\gamma_0(x)$ decreases linearly from
0.7 to 0.57 J/molK$^2$, according to the increase of $T_K$.

A significant contribution to the spectrum of excitations
results from the first excited
crystal field (CF) level, which was estimated as
$\Delta_{CF}^I\approx 25K$ from magnetic susceptibility \cite{8Kita}
and at 33K
from neutron scattering measurements \cite{20CFdoni} on
CePd$_2$Al$_3$. From the latter, the Kondo temperature was evaluated
as $T_K=22K$,
that indicates a peculiar characteristic of this system since
$T_{\rm N}<<T_K$ and $T_K\approx \Delta_{CF}^I$. Further information can be
obtained from specific heat results, where the contribution of the
first excited CF doublet can be observed from measurements up to 30K.
In Fig. 6b we show those results for $x$=0.06 and 0.3. Results
derived for $x$=0.06 are compared with model calculations for a
$\Delta_{CF}^I$=28K splitting and an equal value of $T_K$
\cite{19DesgrRasul}. This relatively high value of $T_K$ does not
match with the value of $\gamma_0$ extracted from the
low temperature specific heat (see Fig. 6a) if it is taken as
$\gamma_0\propto 1/T_K$.
In order to clarify this
aspect we have analyzed the entropy gain ($\Delta S_m$) in those
samples. Following the definition for
 a single Kondo impurity model \cite{21Degr},
$T_K$ is tentatively defined as the temperature where $\Delta S_m \approx$
2/3R$\ln$2. This procedure yields $T_K$ values ranging between 6.3K for
$x$=0.03 and 15K for $x$=0.3. Differences of $T_K$, as derived from
measurements below or above 10K, evidence the relevant role of the first
excited CF level for the high temperature range. Hence, one has to
distinguish between
$T_K$ associated with the GS only $T_K^{GS}$, and $T_K^{CF}$ that
also includes the hybridization effects from the excited level.

\begin{figure}
\begin{center}
\includegraphics[angle=270,width=.47\textwidth] {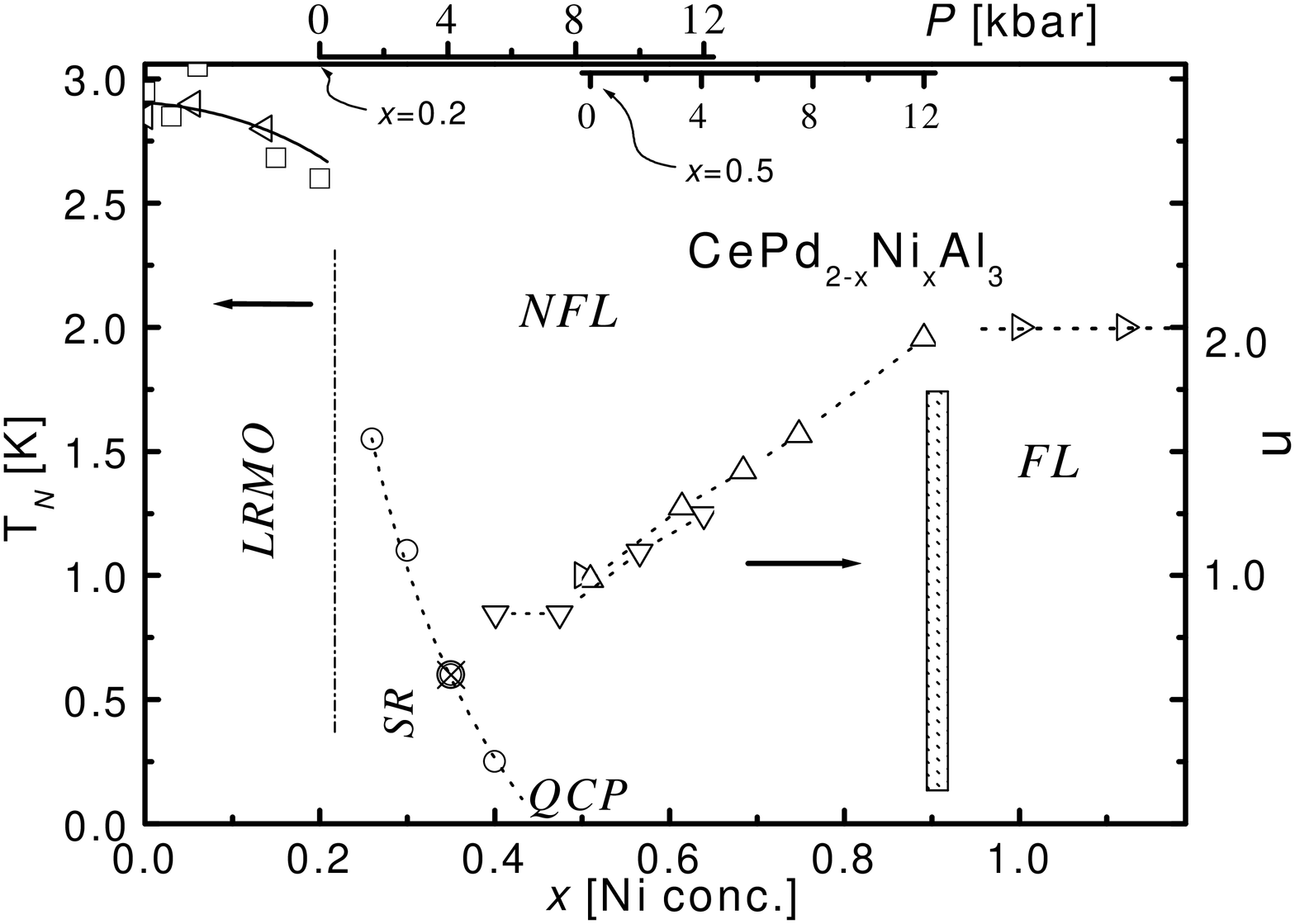}
\end{center}
\caption{Magnetic phase diagram of CePd$_{2-x}$Ni$_x$Al$_3$,
with $T_{\rm N}$ determined from $C_{\rm m}/T$: $\Box$ and $\bigcirc$, and ac-
$\chi$: $\otimes$, in
comparison with pressure effect on $x$=0 ($\lhd$) \cite{14Tang}. Evolution of
exponent $n$ under pressure applied on samples with
$x$=0.2 ($\bigtriangledown$); 0.5 ($\bigtriangleup$) \cite{10Galanta}
and 1.0 ($\rhd$) determined from $R(T)$ measurements.
Lines are guides for the eyes.}
\label{F7PhasDiag}
\end{figure}

Because of the fragility of the samples investigated, the actual
geometrical factor (length/section=$l/s$) cannot be properly
evaluated. Thus, instead of the resistivity $\rho$, only the
resistance $R$ is accessible. Its temperature dependence can be
expressed as:
\begin{equation}
R(T)=R_{\rm 0}+R_{\rm m}(T)+R_{\rm {ph}}(T)
\label{eqn1}
\end{equation}
where $R_{\rm 0}$ is residual resistance,
$R_{\rm m}(T)$ the resistance due to scattering of conduction
electrons with magnetic moments and $R_{\rm {ph}}$ with phonons.

\begin{figure}
\begin{center}
\includegraphics[angle=270,width=.47\textwidth] {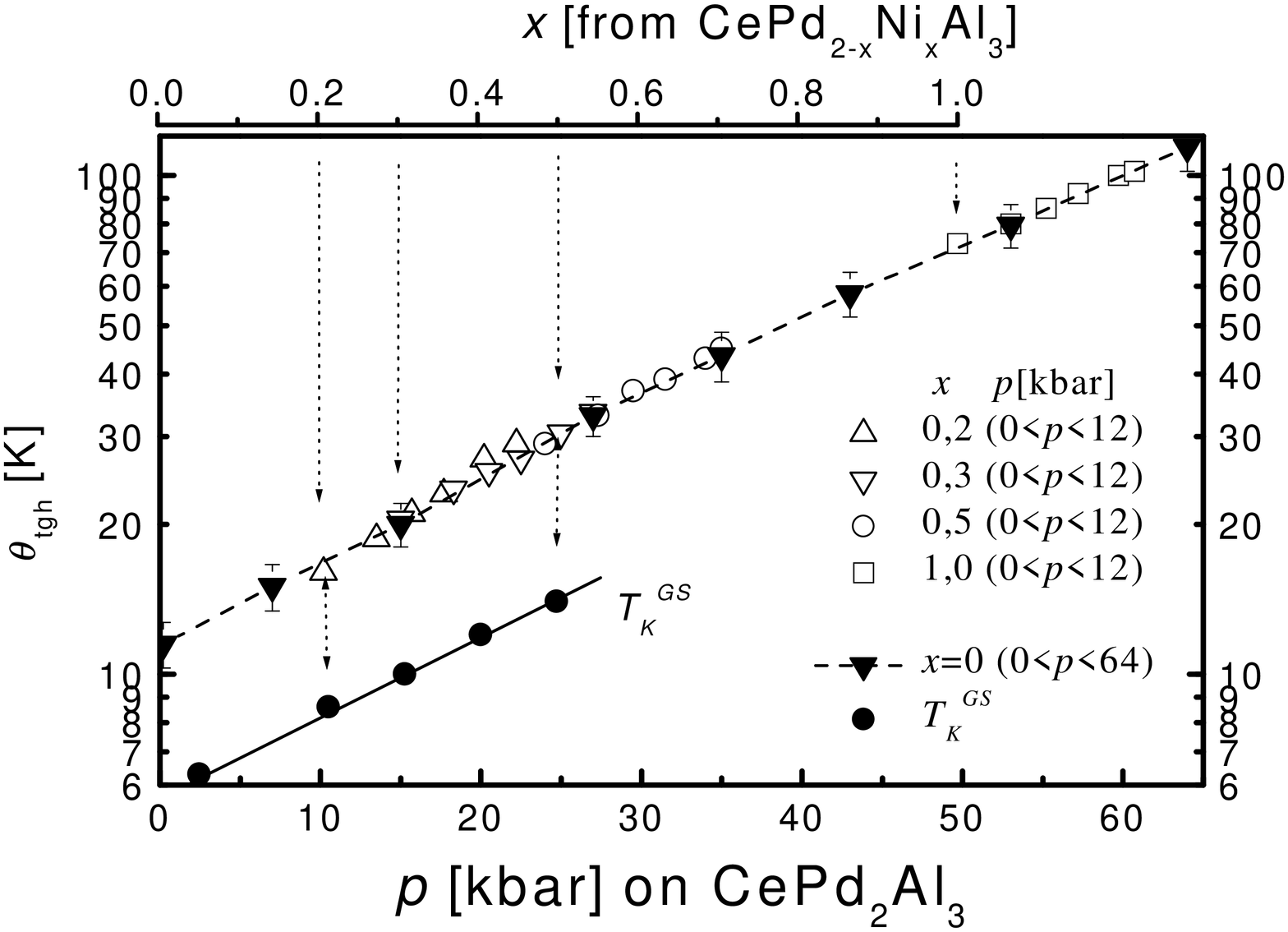}
\end{center}
\caption{Pressure (lower abscissa) and concentration (upper abscissa)
evolution of $\theta_{\rm {tgh}}$ in a semi-logarithmic (ordinate)
representation.
The dotted curve corresponds to a dln$\theta_{\rm {tgh}}$/dx=2.3/Ni at.
Some $T_K^{GS}(x)$ values are also included (see the text). Representative
error bars are
shown on the $x=0$ ($0<p<64kbar$) points) }
\label{F8NewTKs}
\end{figure}

In order to relate the results obtained as a function of concentration and
pressure, one has to look for a unique parameter that allows to compare those
results. $T^R_{\rm {max}}(x,p)$ could be considered for that purpose.
This maximum, however, is the result of a competition between two different
scattering regimes: the incoherent Kondo scattering at high temperature
($T>>T^R_{\rm {max}}$)
associated with the first excited CF level and
the coherent one at $T<T^R_{\rm {max}}$, which arises from the
Ce-Kondo lattice behavior at low temperature.
Since the formation of $\rho_{\rm {max}}$ corresponds
to a continuous change of regime, it is intrinsically broad and (as mentioned
before) it becomes even wider when it moves to high temperature (see Fig.5),
making its evaluation quite speculative above 200K. Another problem concerns
the {\it extensive} character of the terms used in Eqn.\ref{eqn1},
that depend on the
size of the sample. To get a unique and {\it intensive} scaling parameter, one
has to look for a quantity "$\theta$" characterizing the evolution of $R(T)$
in a
large range of temperature, i.e.
\begin{equation}
R(T)= l/s[\rho_0+\rho(T/\theta)+\rho_{\rm {ph}}(T)]
\end{equation}
Following the Gr\"uneisen criterion, we have chosen an heuristic
expression for $R(T)$ which is properly described by a simple
function like:

\begin{equation}
R(T)=R_0+ L\times \tanh(T/\theta_{\rm {tgh}})+R_{ph}(T)
\end{equation}
\noindent where $\theta_{\rm {tgh}}$ is the {\it intensive} parameter which
scales the
measured $R_{\rm m}(T)$ curves and allows to compare all the resistivity
measurements, and L is an extensive free coefficient. $R_{\rm {ph}}(T)$ can be
evaluated from the high
temperature slope of $R(T)$, as shown in Fig.5 in comparison to
the pure phonon contribution of LaPd$_2$Al$_3$.
Scattering of electrons by phonons can be considered nearly independent
of Ni doping or pressure and, additionally, is insignificant at low
temperature.
Examples of fits obtained by applying [eq.3] to experimental
data are shown in Fig. 5, while in Fig. \ref{F8NewTKs} we have collected all
the
$\theta_{\rm {tgh}}$ values extracted from $R_{meas}(T)$ performed at
different Ni concentrations and pressures (including stoichiometric
CePd$_2$Al$_3$ \cite{13Hauser} under pressures up to 64kbar).

The $\theta_{\rm {tgh}}(x)$ dependence, depicted in Fig.8, resembles the
expression for the binding energy of a Kondo singlet T$_K\propto
exp(-1/|J_{ex}N_F|)$ \cite{6Doni}. The relative change of
$\theta_{\rm {tgh}}$ with Ni concentration is: $dln\theta_{\rm {tgh}}/dx$=2.3/
Ni
at. Similar comparison can be done with the pressure dependence,
getting a ratio: $dln\theta_{\rm {tgh}}/dp$=0.042/kbar. Knowing from Ref.
\cite{14Tang} the bulk modulus of CePd$_2$Al$_3$, $B_0=680$~kbar,
one can evaluate an
electronic Gr\"uneisen parameter \cite{22Thompson}
as: $\Omega_e=dln\theta_{\rm {tgh}}/dlnV$=28.5. If the same calculation
is done taking into account the volume change caused by alloying (see
Fig. 1a) one obtains a 2.5 times larger value for $\Omega_e$. This
indicates that "chemical pressure" produces a stronger increase of
$\theta_{\rm {tgh}}$ than hydrostatic pressure, probably because of
stronger modifications of the Fermi energy. In order to make a
quantitative comparison between $\theta_{\rm {tgh}}(x)$ and the Kondo
temperature, we have included in Fig. 8 the $T_K^{GS}$ values
extracted from the evolution of $\Delta S_m(T)$. We find that between
$x$=0 and 0.5, $\theta_{\rm {tgh}}(x)\approx 2 T_K^{GS}(x)$.

\section{Conclusions}

The magnetic phase diagram of the Ce series investigated in this
work is identified as belonging to the class of phase boundaries
where $T_{\rm N}$ is weakly pressure or doping dependent before
vanishing (i.e. group iii). From an accurate investigation of the
critical concentration region, we have confirmed that stable
AF-LRMO collapses in CePd$_{2-x}$Ni$_x$Al$_3$  at $x$=0.2.
However, above that concentration a remainder SR interaction,
probably caused by increasing atomic disorder, can be recognized.
Between the collapse of LRMO and the critical point a coexistence
of two contributions was identified. One corresponds to a
decreasing SR order component that vanishes at the QCP (0.45 $\leq
x_{cr}\leq$ 0.5), and the other to a NFL contribution. The latter
dominates the low temperature entropy and is well accounted for by
the function $C_{\rm {NFL}}/T=\gamma_0 (1-\sqrt{T/D})$, with a
constant factor D $\approx$18K. The decrease of $\gamma_0$ between
$x=0.3$ and 0.45 might originate in the increase of $T_K^{GS}$ in
that range of concentration. Above the critical point the system
keeps its NFL behavior up to $x=1$, where the FL phase sets in.
This indicates that the loss of long or short range magnetic order
does not necessarily imply an immediate crossover to a FL
behavior. Because of the proximity of the first  excited CF level
to the GS, there is a substantial CF contribution to the physical
properties of the system. This requires to make a clear
distinction between $T_K^{GS}$ evaluated at low temperature and
$T_K^{CF}$ extracted from measurements performed above about 10K.

Doping and pressure effects were successfully compared by introducing a
scaling parameter proportional to the Kondo temperature. This comparison
corroborates
that the effect of both control parameters is not identical because
chemical pressure produced by doping may not only affect the volume but also
the
Fermi energy and details of the DOS of the system.

\section*{Acknowledgments}
We are grateful to M.G. Berisso for his experimental support in
$\chi_{ac}$ measurements. J.G.S. and P.P. are Conicet fellows.
This work was partially supported by the FWF-Conicet International
Cooperation Program 097/02 and Austrian FWF P16370 and a Grant-in-Aid for Scientific Research from MEXT of Japan.

\end{document}